%% file: main.tex
\documentclass{article}

\usepackage[final]{neurips_2021}
\usepackage{graphicx}

\usepackage[utf8]{inputenc} 
\usepackage[T1]{fontenc}    
\usepackage{hyperref}       
\usepackage{url}            
\usepackage{booktabs}       
\usepackage{amsfonts}       
\usepackage{nicefrac}       
\usepackage{microtype}      
\usepackage{xcolor}         
\usepackage{caption}
\usepackage{subcaption}
\usepackage{wrapfig}

\input{commands.tex}

\usepackage{natbib} 
\setcitestyle{square,numbers,comma}
\makeatletter
\newcommand{\printfnsymbol}[1]{%
  \textsuperscript{\@fnsymbol{#1}}%
}
\makeatother

\title{Using Non-Linear Causal Models to Study Aerosol-Cloud Interactions in the Southeast Pacific}

\author{%
  Andrew Jesson\thanks{Equal contribution.}\\
  OATML\\
  Department of Computer Science\\
  University of Oxford \\
  \texttt{andrew.jesson@cs.ox.ac.uk} \\
  \And
  Peter Manshausen\printfnsymbol{1} \\
  Atmospheric, Oceanic, and Planetary Physics\\
  Department of Physics\\
  University of Oxford \\
  \texttt{peter.manshausen@physics.ox.ac.uk} \\
  \AND
  Alyson Douglas\printfnsymbol{1} \\
  Atmospheric, Oceanic, and Planetary Physics\\
  Department of Physics\\
  University of Oxford \\
  \texttt{alyson.douglas@physics.ox.ac.uk} \\
   \And
  Duncan Watson-Parris \\
  Atmospheric, Oceanic, and Planetary Physics\\
  Department of Physics\\
  University of Oxford \\
  \texttt{duncan.watson-parris@physics.ox.ac.uk} \\
  \And
  Yarin Gal \\
  OATML \\
  Department of Computer Science\\
  University of Oxford \\
  \texttt{yarin@cs.ox.ac.uk} \\
  \And
  Philip Stier \\
  Atmospheric, Oceanic, and Planetary Physics\\
  Department of Physics\\
  University of Oxford \\
  \texttt{philip.stier@physics.ox.ac.uk} \\
}

\begin{document}

\maketitle

\begin{abstract}
Aerosol-cloud interactions include a myriad of effects that all begin when aerosol enters a cloud and acts as cloud condensation nuclei (CCN). 
An increase in CCN results in a decrease in the mean cloud droplet size (r$_{e}$).
The smaller droplet size leads to brighter, more expansive, and longer lasting clouds that reflect more incoming sunlight, thus cooling the earth.
Globally, aerosol-cloud interactions cool the Earth, however the strength of the effect is heterogeneous over different meteorological regimes.
Understanding how aerosol-cloud interactions evolve as a function of the local environment can help us better understand sources of error in our Earth system models, which currently fail to reproduce the observed relationships. 
In this work we use recent non-linear, causal machine learning methods to study the heterogeneous effects of aerosols on cloud droplet radius.
\end{abstract}

\newpage
\section{Clouds remain  the largest source of uncertainty for future climate projections}

\begin{wrapfigure}{r}{0.5\textwidth}
    \begin{center}
        \includegraphics[width=0.5\textwidth]{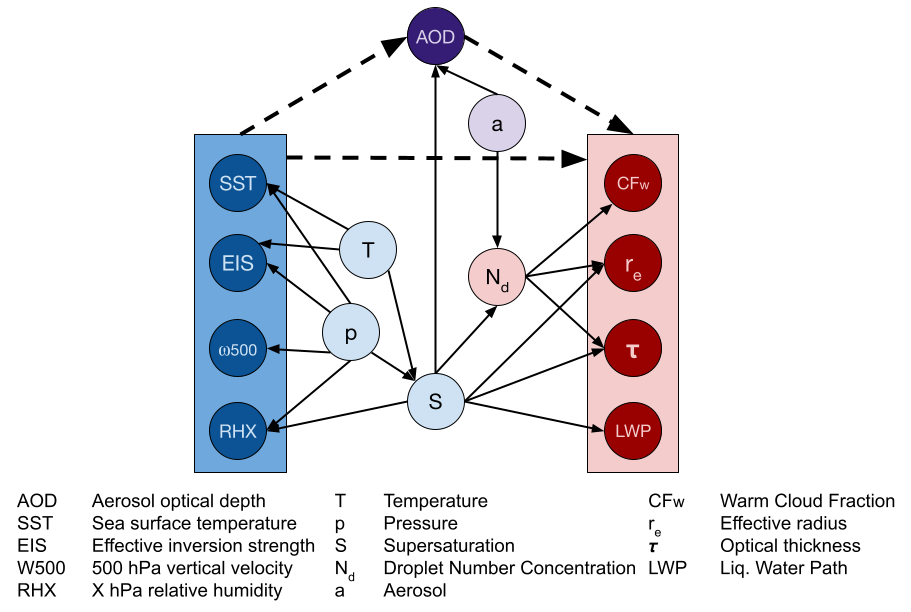}
    \end{center}
    \caption{
        Causal diagram of the aerosol and aerosol proxy AOD (top, purple), affected cloud properties (red, right), and the environmental confounders (blue).
    }
    \label{fig:causal_diagram}
\end{wrapfigure}

Aerosol-cloud interactions include a myriad of effects that are initiated when aerosol, released through natural or anthropogenic activities, enters a cloud and acts as cloud condensation nuclei (CCN). 
Theoretically, if you hold the liquid water content of a cloud constant, an increase in CCN results in a decrease in the mean cloud droplet size (r\(_{e}\)).
These smaller droplets increase the brightness of the cloud \cite{twomey1977influence} and delay precipitation formation \cite{albrecht1989aerosols}.
The resulting brighter, larger, and longer lasting cloud reflects more incoming solar radiation thus cooling the earth. 
The IPCC 6th report estimates an increase in the magnitude of cooling due to aerosol-cloud interactions compared to the 5th report, without any improvement to the confidence level \cite{ipcc6ts}.
Aerosol-cloud interactions, and in particular the adjustments due to the delay in precipitation on cloud amount and liquid water content, remain a large source of uncertainty in future climate projections of global warming. 

Globally, aerosol-cloud interactions work to cool the Earth, however the strength of the effect is heterogeneous. One source of heterogeneity is modulation by local meteorology \cite{ackerman2004impact, small2009can, chen2014satellite, douglas2020quantifying}. Understanding how the local environment influences aerosol-cloud interactions can help us improve our Earth system model parameterizations of these effects and outcomes. In this work we use recent non-linear causal machine learning methods \cite{athey2019estimating, jesson2021quantifying} to study the heterogeneous effects of aerosols on cloud droplet radius.

\section{Background}

\subsection{Meteorological Influence on Aerosol-Cloud Interactions}

Aerosol-cloud interactions can be labelled as a heterogeneous effect, as their sign and magnitude is modulated by a cloud's local meteorology. 
The local meteorology confounds the estimated effect size as it impacts both the properties of the cloud as well as the aerosol conditions in the cloud's environment. The relationships between the local meteorology, cloud properties, and aerosol are illustrated in Figure \ref{fig:causal_diagram}. 
Cloud properties (in red) are dependent on the amount of suitable aerosol (a, light purple) that potentially can be activated as cloud condensation nuclei (increasing Nd, light red); activation itself is dependent on the level of supersaturation within a cloud and the size, shape, and type of aerosol (S, light blue). 
Supersaturation itself cannot be directly observed using satellites and is dependent on other similarly unobserved meteorological variables like the temperature and pressure profiles within a cloud \cite{kohler1936nucleus}. 
For these hidden variables (T, p, and S), we can rely on indirect diagnostics such as estimated inversion strength (EIS), sea surface temperature (SST), upper level convergence/divergence (w500), and relative humidities at 700 and 850 mb (RHx) (dark blue) to approximate their effects on aerosol (a). 

The dependence on indirect diagnostics underscores how evaluating relationships between aerosol and observed cloud properties, without accounting for confounders, is fundamentally flawed. 
Furthermore, we do not have a direct way of observing the amount of aerosol in the atmosphere or the number of cloud condensation nuclei within a cloud, instead relying on aerosol optical depth (AOD), an indirect measure of the amount of aerosol in a column of atmosphere. 
However, aerosol optical depth is subject to hygroscopic growth, or swelling in the presence of high humidity environments such as near cloud edges \cite{christensen2017unveiling}, an additional confounding influence on the effect of aerosol on cloud properties. 

\begin{table}
  \caption{Satellite Observations and Climate Reanalysis Used}
  \label{tab:obstable}
  \centering
  \begin{tabular}{lc}
    \toprule
    \multicolumn{2}{c}{}                   \\
    Product name     & Description \\
    \midrule
    Droplet Radius (r\(_{e}\)) & MODIS 1.6, 2.1, and 3.7 \(\mu\)m channels    \cite{baum2006introduction} \\
    Precipitation & NOAA CMORPH Climate Data Record \cite{prat2021global}\\
    Sea Surface Temperature & NOAA WHOI Climate Data Record \cite{cogan1976measurement}\\
    Vertical Motion at 500 mb & MERRA-2 Reanalysis \cite{bosilovich2015merra}\\
    Estimated Inversion Strength &  MERRA-2 Reanalysis \cite{wood2006relationship, gelaro2017modern}\\
    Relative Humidity\(_{700}\) \& RH\(_{850}\) & MERRA-2 Reanalysis \cite{gelaro2017modern}\\
    Aerosol Optical Depth (AOD) & MERRA-2 Reanalysis \cite{gelaro2017modern}\\
    \bottomrule
  \end{tabular}
\end{table}

\subsection{Causal Machine Learning for Heterogeneous Effect Estimation}

The potential outcomes framework of causal inference \citep{neyman1923applications, rubin1974estimating} provides a principled methodology for estimating the heterogeneous the effect of a binary treatment $\T \in \{0, 1\}$ on outcomes $\Y$ for units described by covariates $\X$. 
The treatment effect for a unit $u$ is defined as the difference in potential outcomes $\Yone(u) - \Yzero(u)$, where the r.v. $\Yone$ represents the potential outcome were the unit \emph{treated}, and the r.v. $\Yzero$ represents the potential outcome were the \emph{not treated}.
Realizations of the random variables $\X$, $\T$, $\Y$, $\Yzero$, and $\Yone$ are denoted by $\x$, $\tf$, $\y$, $\yzero$, and $\yone$, respectively.

The unit level treatment effect is a fundamentally unidentifiable quantity, so instead we look at the Conditional Average Treatment Effect (CATE): $\mathrm{CATE}(\x) \equiv \E[\Yone - \Yzero \mid \X = \x]$ \cite{abrevaya2015estimating} to estimate heterogeneous effects.
The CATE is identifiable from an observational dataset $\D = \left\{ (\x_i, \tf_i, \y_i)\right\}_{i=1}^n$ under the following three assumptions: \textbf{Consistency} $\y = \tf \yt + (1 - \tf) \y^{1 - \tf}$, i.e. an individual's observed outcome $\y$ given assigned treatment $\tf$ is identical to their potential outcome $\yt$; \textbf{Unconfoundedness} $(\Yzero, \Yone) \indep \T \mid \X$.; \textbf{Overlap} $0 < \pi_{\tf}(\x) < 1: \forall \tf \in \mathcal{T}$, where $\pi_{\tf}(\x) \equiv \mathrm{P}(\T = \tf \mid \X = \x)$ is the \emph{propensity for treatment} for individuals described by covariates $\X = \x$ \cite{rubin1974estimating}.
When these assumptions are satisfied, $\widehat{\mathrm{CATE}}(\x) \equiv \E[Y \mid \T = 1, \X = \x] - \E[Y \mid \T = 0, \X = \x]$ is an identifiable, unbiased estimator of $\mathrm{CATE}(\x)$. The population level average treatment effect (ATE) is then the expectation of the CATE over all $\X$.

When using satellite observations as we are within (presented in Table \ref{tab:obstable}), it is common for these assumptions to fail. Nonetheless, these assumptions give us a perspective to understand why we should be uncertain about the effects we estimate from observational data \cite{jesson2020identifying, jesson2021quantifying}. We use the scalable, uncertainty-aware machine learning methodology of \citet{jesson2021quantifying} to model non-linear, heterogeneous aerosol-cloud interactions.  

\section{Exploring Heterogeneous Aerosol Cloud Interactions using Non-linear Causal Models}

First, we need to translate the problem of estimating aerosol-cloud interactions into a causal problem.
We consider SST, EIS, RH$_{700}$, RH$_{850}$, and w500 as our covariates $\X$. The AOD is our treatment variable $\T$.  We discretize the measured AOD by applying a threshold at its median value of 0.3. Examples with raw AOD values less than 0.07 and greater than 1.0 are discarded. Finally, we look at each of $r_e$, $CF_w$, $\tau$, and LWP as our outcome variables $\Y$, but focus our analysis on $r_e$ in the main text. We further focus on non-precipitating clouds by discarding examples with precipitation greater than 0.05. We look at daily averages between 2003 and 2020. We generate training, validation, and test sets by partitioning weekdays into the training set, Saturdays into the validation set, and Sundays into the test set.

We use Quince \cite{jesson2021quantifying} to estimate the heterogeneous effects, which are validated against estimates from Bayesian linear regression \cite{mackay1992bayesian, tipping2001sparse} and causal forest \cite{athey2019estimating} in the appendix. Implementation details of each are given in the appendix.  

\begin{figure}
  \centering
      \begin{subfigure}[b]{14cm}
          \centering
          \includegraphics[width=\textwidth]{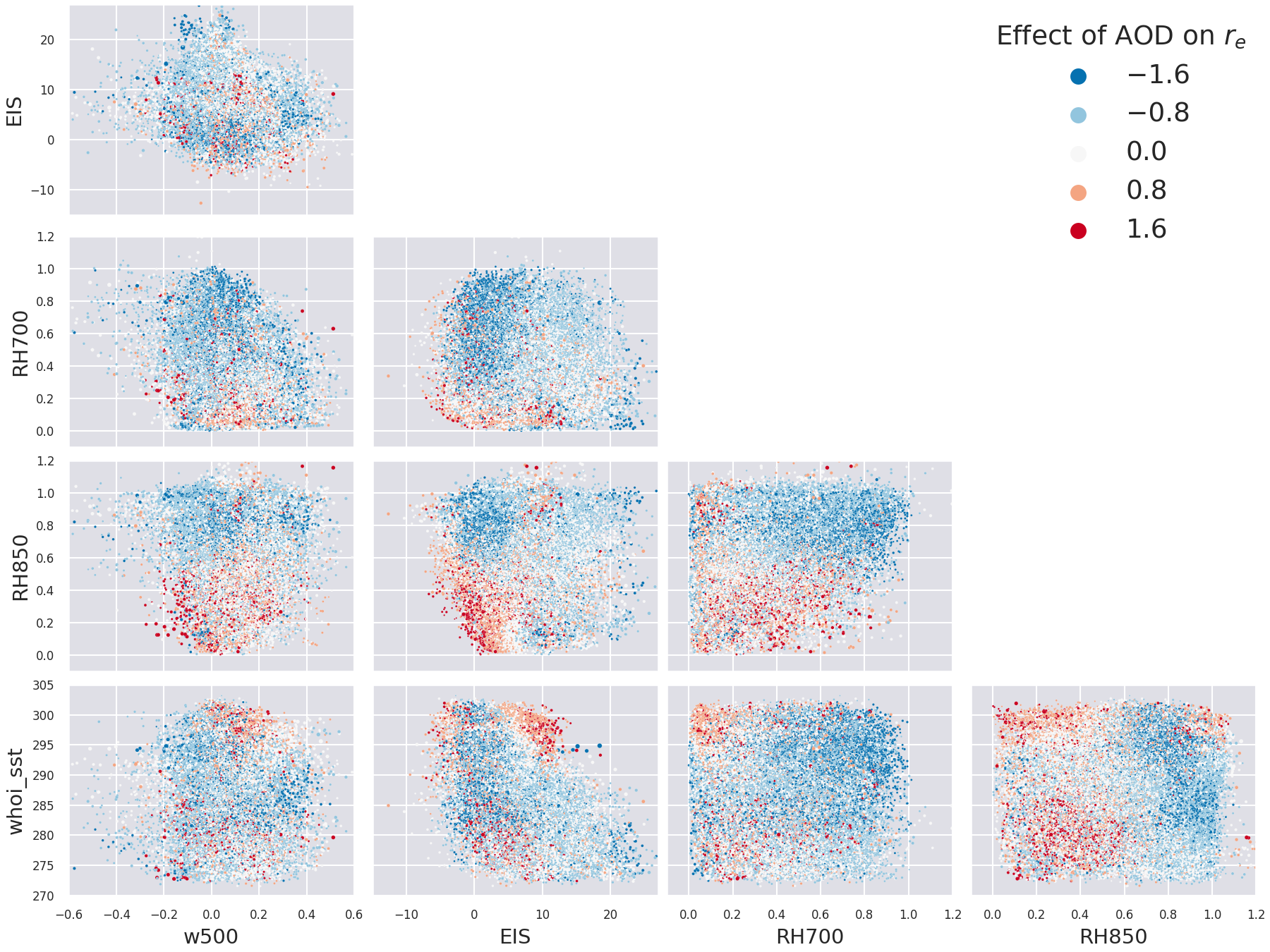}
          \caption{}
          \label{fig:heatmaps}
      \end{subfigure}
        \begin{subfigure}[b]{14cm}
        \centering
        \includegraphics[width=\textwidth]{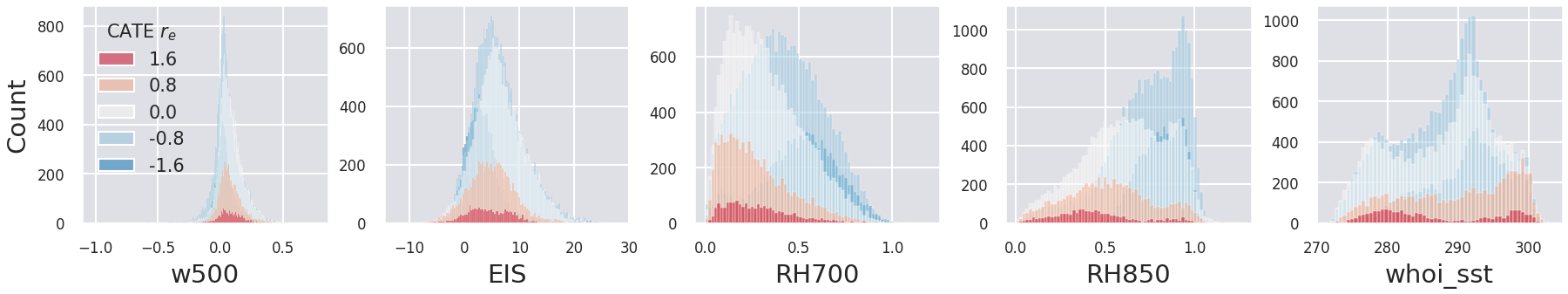}
        \caption{}
        \label{fig:hist_quince_lre}
    \end{subfigure}
     \caption{The effect estimates (a) of AOD on r\textsubscript{e} using Quince for each of the ten possible environmental combinations and the histograms (b) of these effects against each of the environmental parameters. }
     
     \label{fig:heterogeneous}
\end{figure}

\section{Results}
The average treatment effect (ATE) for AOD on r\textsubscript{e}, measured as the difference in r\textsubscript{e} between above average and below average aerosol cases, varies from -.359 \(\pm\).533 to -.452 \(\pm\).010 depending on the causal method chosen (Table \ref{tab:ate_lre}). On average, an increase in AOD results in a reduction of r\textsubscript{e}. Positive effects of AOD on r\textsubscript{e} have been found to occur under certain conditions, but as expected are rare occurrences \cite{yuan2008increase, jiang2006aerosol}. The heterogeneous effects of AOD on r\textsubscript{e} from the Quince model, or the impact of different meteorological conditions on the effect's magnitude and sign, are shown in Figure \ref{fig:heterogeneous}. The effect's magnitude and sign are directly impacted by changing environmental regime, demonstrating that aerosol effects on r\textsubscript{e} are an example of a heterogeneous effect. By evaluating the effect as a function of different environmental parameters, the processes responsible for altering the sign or magnitude can be better understood. For example, in the RH850 vs. EIS diagram, the effect goes from negative (decreasing r\textsubscript{e} with increasing AOD) to positive (increasing r\textsubscript{e} with increasing AOD) as RH850 and stability both decrease. In unstable, dry conditions, convection can occur, so the effect of aerosol is dependent on the humidity transported into the cloud layer \cite{douglas2021global}. Conversely, in humid and stable conditions, clouds are prone to be aerosol-limited, meaning any increase in aerosol leads to a large, negative impact on the r\textsubscript{e} \cite{koren2014aerosol}.

In order to employ Quince to quantify aerosol-cloud interactions using satellite observations, a number of assumptions are made that may alter the overall quality of the predictions. The first being that aerosol optical depth directly impacts r\textsubscript{e}, which we know is not the case. AOD is a proxy for aerosol and is not a direct measure of the amount of aerosol present in the atmosphere. We did not include how aerosols, clouds, and confounding effects are spatially correlated. Any spatial correlations between the environment, aerosol, and the effect size is ignored by Quince, but may be acting as a hidden confounding influence. Additionally, given we are only using proxies (EIS, SST, w500, and RHX) of the true confounders (T, p, and S), there are likely to exist other environmental features that additionally capture the true confounders influence on the effect size of aerosol on r\textsubscript{e}. Currently, we are using daily averages at a 1° x 1° scale, which may be restricting the predictive power of our predictions, as the scale of the interactions is on the order of kilometers (cloud scale), which may not be captured by the regional, daily mean. 

We do not have a true, counterfactual value to compare our predictions against, making it difficult to validate either model as the “best” model for evaluating the causal, heterogeneous relationships of aerosol-cloud interactions. In theory, Quince should better resolve the effect size as it incorporates feature extraction and inherent relationship between feature correlations with both the predictors (r\textsubscript{e}) and treatment (AOD), however this is difficult to definitively state. Quince does show a reduced range of uncertainty compared to other methods that allow for environmental causal attribution on effect size, like a Bayesian linear regressions or causal forests (Table \ref{tab:ate_lre}).

\begin{table}
  \caption{Comparison of the average effect size estimates for three different causal models. The mean and standard deviation of the approximate posterior distribution (BLR) or mean and standard error of the model ensemble (Causal Forest and Quince) are reported along with the test set R$^{2}$ values for regressing the outcome given the observed AOD values.}
  \label{tab:ate_lre}
  \centering
  \begin{tabular}{lcccc}
    \toprule
    & \multicolumn{3}{c}{Estimated ATE of above vs. below average AOD on $r_e$} & R$^{2}$ \\
    Method & Train & Valid & Test & Test \\
    \midrule
    BLR & $-.287\pm.015$ &  &  & $.15\pm$ \\ 
    Causal Forest & $-.328\pm.568$ & $-.325\pm.569$ & $-.359\pm.533$ & $.21\pm$ \\
    Quince & $-.429\pm.009$ & $-.413\pm.010$ & $-.452\pm.010$ & $.23\pm$ \\
    \bottomrule
  \end{tabular}
\end{table}

\vspace{-1em}
\section{Conclusions}
\vspace{-1em}

Herein we show how the heterogeneous effects of aerosol on r\textsubscript{e} can be evaluated using Quince, a causal neural network. This effect, and other aerosol-cloud interactions, are primed to be untangled using causal methods; non-linear, causal models such as causal forests or Quince are the ideal tools to evaluate the complex interactions between aerosol, clouds, and the environment. These models can be exploited to flexibly model the causal relationships between treatments (aerosol loading), covariates (local meteorological influence), and outcomes (changes in cloud properties). Aerosol-cloud interactions are a high-impact science problem for new causal inference methods.

\section{Acknowledgements}
This project was supported by the European Union’s Horizon 2020 research and innovation program under grant agreement No 821205 (FORCeS) and Marie Skłodowska-Curie grant agreement No 860100 (iMIRACLI).
PS and AD were supported by the European Research Council (ERC) project constRaining the EffeCts of Aerosols on Precipitation (RECAP) under the European Union’s Horizon 2020 research and innovation programme with grant agreement no. 724602.

\bibliography{references.bib}

\newpage
\appendix

\section{Additional Results}

Table \ref{tab:ate} enumerates the estimated ATE of AOD on each of $CF_w$, $r_e$, $\tau$, and LWP. We see that the non-linear causal models estimate AOD having a larger magnitude average effect on $CF_w$ and $r_e$ in the south pacific compared to the linear model estimate. Conversely we see that the non-linear causal models estimate AOD having a smaller magnitude average effect on $\tau$, and LWP in the south pacific compared to the linear model estimate.

\begin{table}[ht]
  \caption{Comparison of the average treatment effect (ATE) estimates of different methodologies. The mean and standard deviation of the approximate posterior distribution (BLR) or mean and standard error of the model ensemble (Causal Forest and Quince) are reported.}
  \label{tab:ate}
  \centering
  \begin{tabular}{lcccccc}
    \toprule
    & \multicolumn{3}{c}{Estimated ATE of AOD on $CF_w$} & \multicolumn{3}{c}{Estimated ATE of AOD on $r_e$} \\
    Method & Train & Valid & Test & Train & Valid & Test \\
    \midrule
    BLR & $-.002\pm.001$ &  &  & $-.29\pm.02$ &  &  \\
    Forest & $.003\pm.030$ & $.005\pm.030$ & $.004\pm.029$  & $-.33\pm.57$ & $-.33\pm.57$ & $-.36\pm.53$ \\
    Quince & $.008\pm.001$ & $.008\pm.001$ & $.008\pm.001$  & $-.43\pm.01$ & $-.41\pm.01$ & $-.45\pm.01$ \\
    \bottomrule \\
    \toprule
    & \multicolumn{3}{c}{Estimated ATE of AOD on $\tau$} & \multicolumn{3}{c}{Estimated ATE of AOD on LWP} \\
    Method & Train & Valid & Test & Train & Valid & Test \\
    \midrule
    BLR & $2.34\pm0.04$ &  &  & $17.8\pm0.40$ &  &  \\
    Forest & $2.05\pm1.37$ & $2.03\pm1.37$ & $2.02\pm1.27$ & $13.3\pm15.6$ & $13.0\pm15.6$ & $12.8\pm14.5$ \\
    Quince & $1.64\pm0.04$ & $1.61\pm0.04$ & $1.61\pm0.04$ & $8.49\pm0.46$ & $8.34\pm0.47$ & $8.05\pm0.46$ \\
    \bottomrule
  \end{tabular}
\end{table}

\begin{figure}[ht]
     \centering
     \begin{subfigure}[b]{0.95\textwidth}
         \centering
         \includegraphics[width=\textwidth]{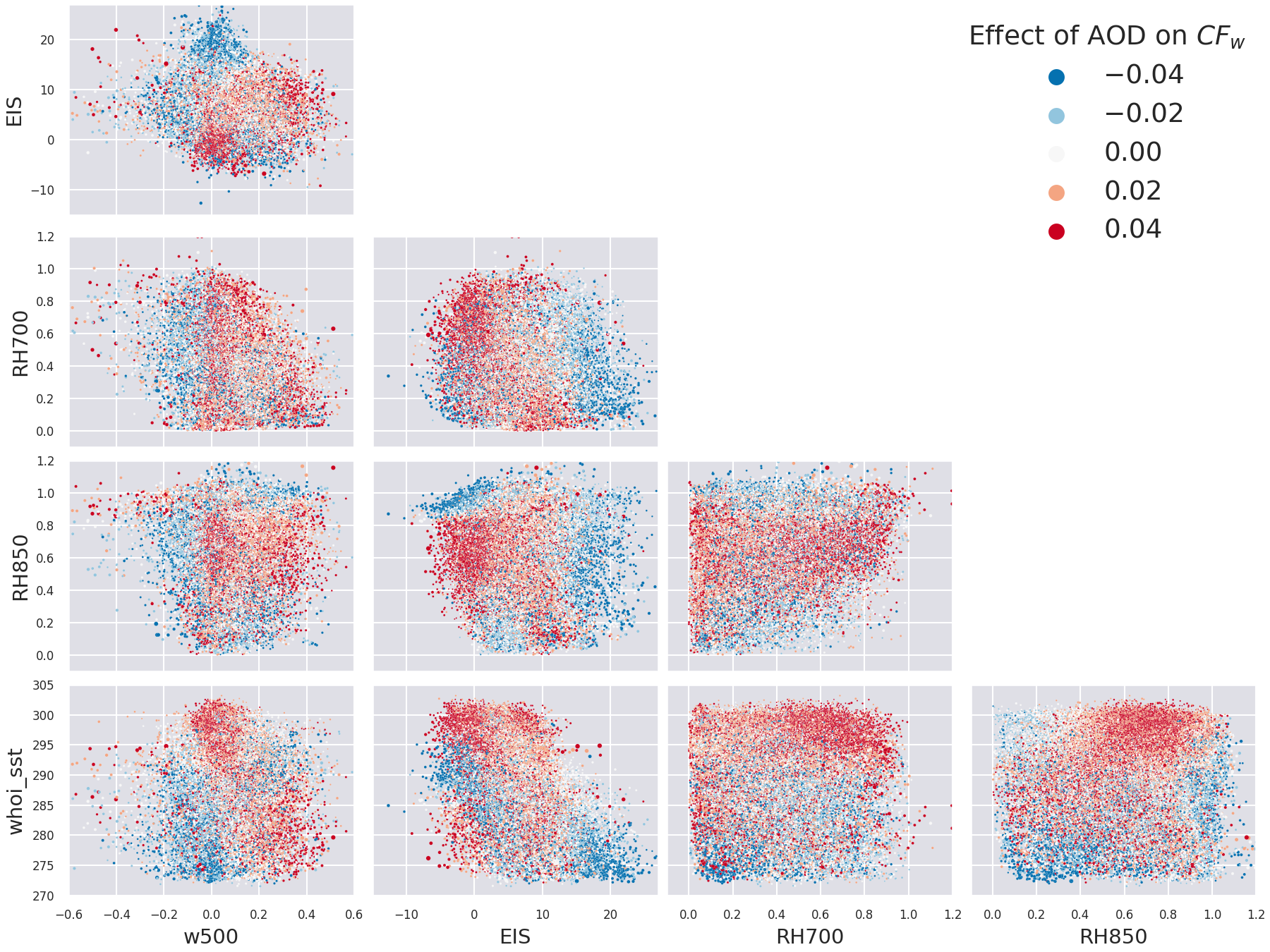}
         \caption{}
         \label{fig:heatmaps_cf}
     \end{subfigure}
     
     \begin{subfigure}[b]{0.95\textwidth}
         \centering
         \includegraphics[width=\textwidth]{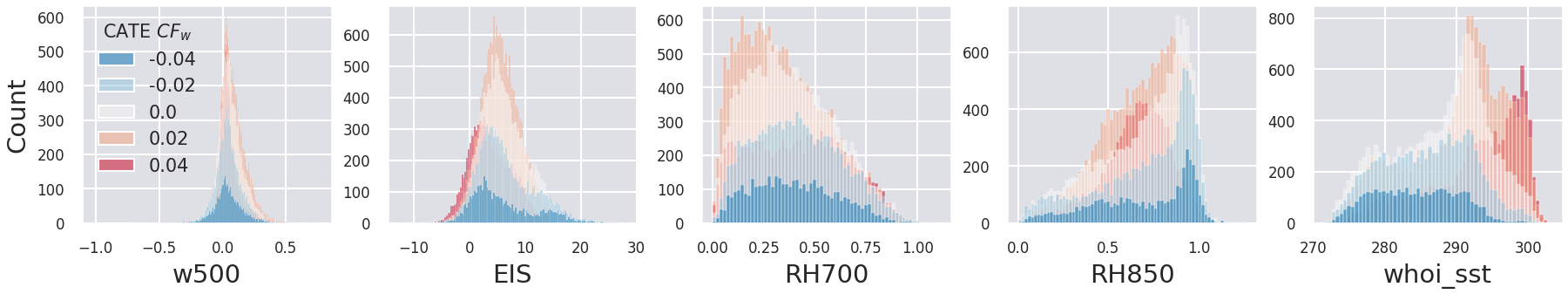}
         \caption{}
         \label{fig:historgrams_cf}
     \end{subfigure}
    \caption{The treatment effect of AOD on $CF_w$, defined as difference between the high and low aerosol. (a) Each plot shows the effect size as the color of scatter points, while the position of the points indicates the values of the observed meteorological covariates. (b) Histograms of effect sizes plotted against each individual meteorological covariate.}
    \label{fig:heterogeneous_cf}
\end{figure}

In Figure \ref{fig:heterogeneous_cf} we look at the heterogeneity of Quince estimated effect sizes of AOD on $CF_w$ across pairs of covariates.

\begin{figure}[ht]
     \centering
     \begin{subfigure}[b]{0.95\textwidth}
         \centering
         \includegraphics[width=\textwidth]{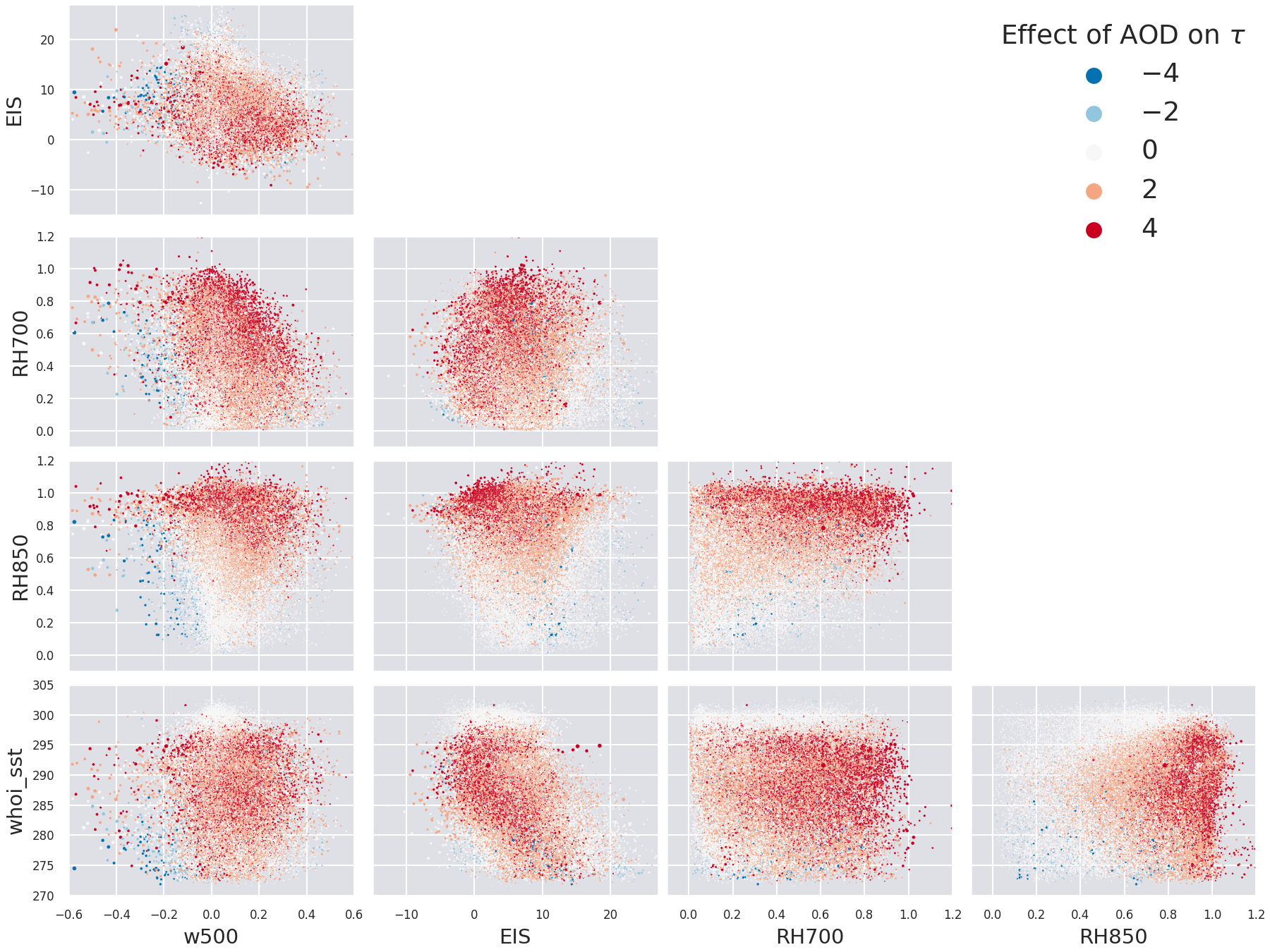}
         \caption{}
         \label{fig:heatmaps_cod}
     \end{subfigure}
     
     \begin{subfigure}[b]{0.95\textwidth}
         \centering
         \includegraphics[width=\textwidth]{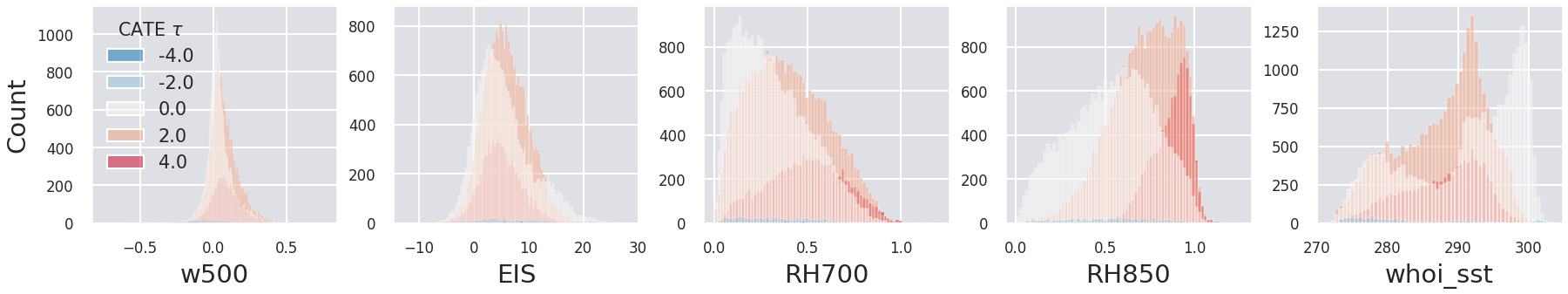}
         \caption{}
         \label{fig:historgrams_cod}
     \end{subfigure}
    \caption{The treatment effect of AOD on $\tau$ (Cloud Optical Thickness), defined as difference between the high and low aerosol. (a) Each plot shows the effect size as the color of scatter points, while the position of the points indicates the values of the observed meteorological covariates. (b) Histograms of effect sizes plotted against each individual meteorological covariate.}
    \label{fig:heterogeneous_cod}
\end{figure}

In Figure \ref{fig:heterogeneous_cod} we look at the heterogeneity of Quince estimated effect sizes of AOD on $\tau$ across pairs of covariates.

\begin{figure}
     \centering
     \begin{subfigure}[b]{0.95\textwidth}
         \centering
         \includegraphics[width=\textwidth]{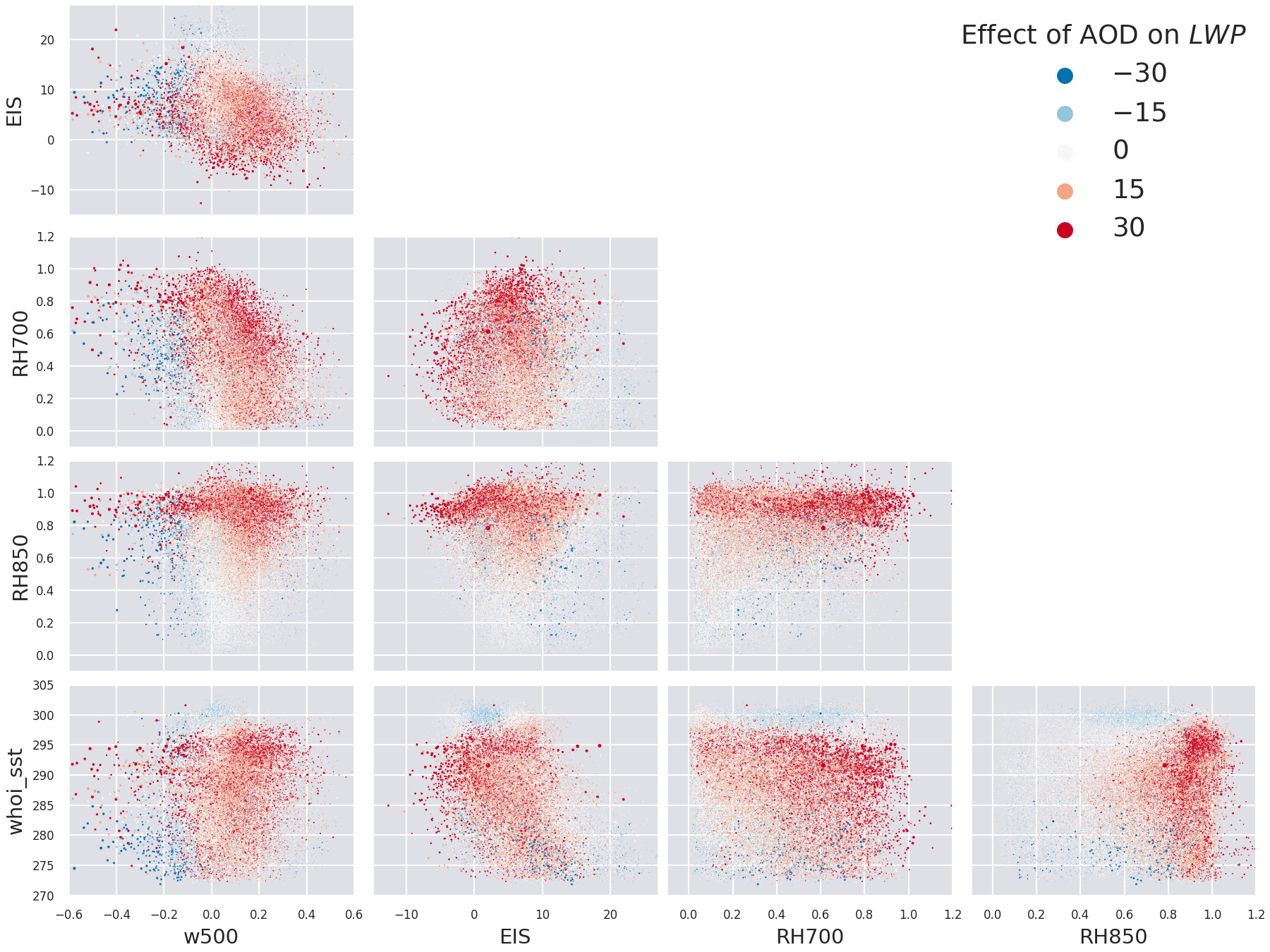}
         \caption{}
         \label{fig:heatmaps_lwp}
     \end{subfigure}
     
     \begin{subfigure}[b]{0.95\textwidth}
         \centering
         \includegraphics[width=\textwidth]{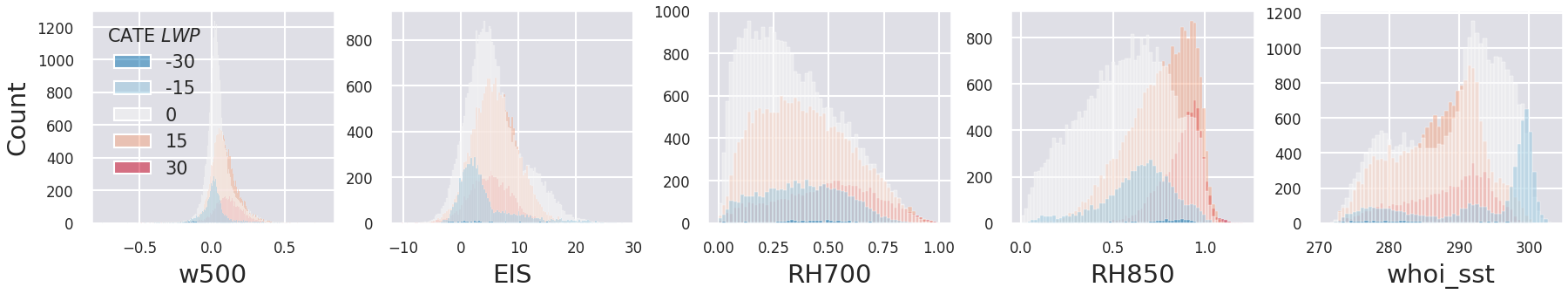}
         \caption{}
         \label{fig:historgrams_lwp}
     \end{subfigure}
    \caption{The treatment effect of AOD on LWP, defined as difference between the high and low aerosol. (a) Each plot shows the effect size as the colour of scatter points, while the position of the points indicates the values of the observed meteorological covariates. (b) Histograms of effect sizes plotted against each individual meteorological covariate.}
    \label{fig:heterogeneous_lwp}
\end{figure}

In Figure \ref{fig:heterogeneous_lwp} we look at the heterogeneity of Quince estimated effect sizes of AOD on LWP across pairs of covariates.

\section{Sanity Checks}

\begin{figure}[ht]
     \centering
     \begin{subfigure}[b]{0.95\textwidth}
         \centering
         \includegraphics[width=\textwidth]{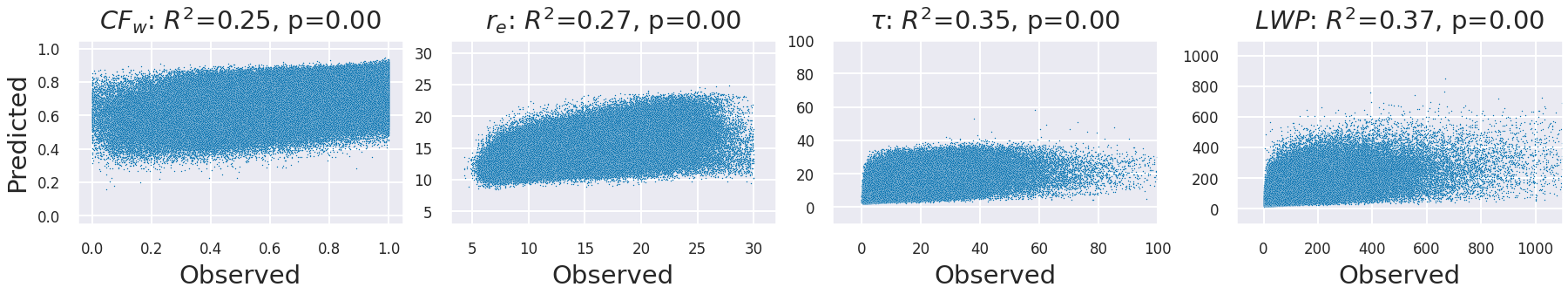}
         \caption{Train}
         \label{fig:scatter_train}
     \end{subfigure}
     
     \begin{subfigure}[b]{0.95\textwidth}
         \centering
         \includegraphics[width=\textwidth]{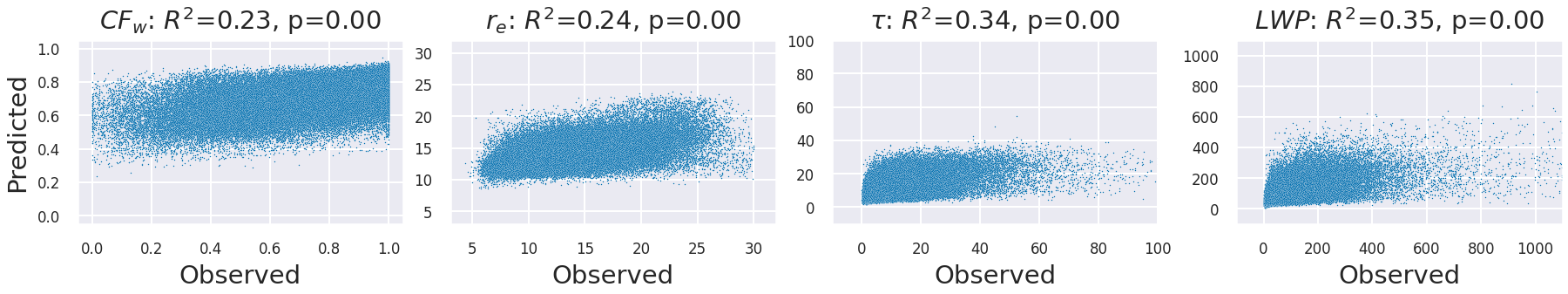}
         \caption{Validation}
         \label{fig:scatter_valid}
     \end{subfigure}
     
     \begin{subfigure}[b]{0.95\textwidth}
         \centering
         \includegraphics[width=\textwidth]{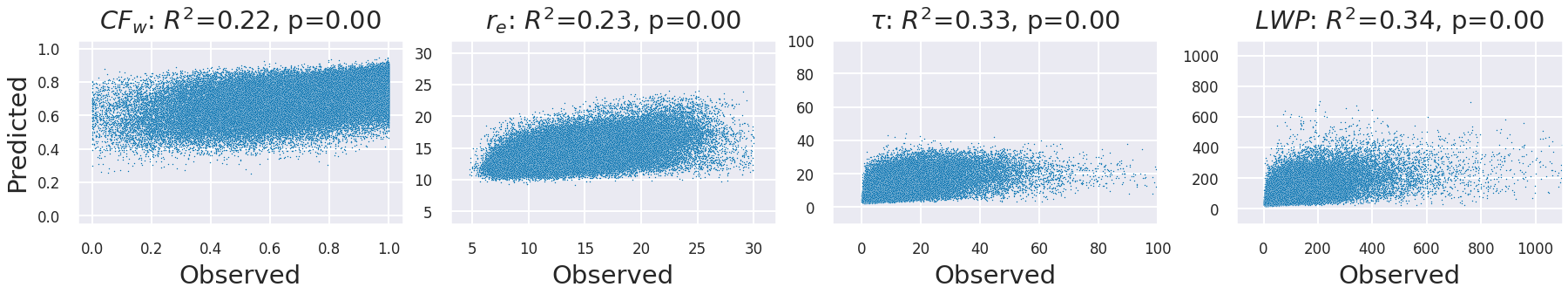}
         \caption{Test}
         \label{fig:fig:scatter_test}
     \end{subfigure}
    \caption{Comparing regression accuracy of Quince on train, validation and test splits. The squared Pearson R coefficient is shown with associated p-value. We can see consistent performance across all splits providing evidence that our results will generalize.}
    \label{fig:scatter_comparison}
\end{figure}

It is impossible to observe ground truth treatment effects, so validating the truth of the reported results is complicated. As a first measure of sanity, we look at the accuracy of regressing the outcome, measured by the coefficient of determination (R$^2$) between predicted and observed outcomes for each of $CF_w$, $r_e$, $\tau$, and LWP. In Figure \ref{fig:scatter_comparison}, we show these results for Quince across the train, validation, and test data splits. 

\begin{figure}[ht]
     \centering
     \begin{subfigure}[b]{0.95\textwidth}
         \centering
         \includegraphics[width=\textwidth]{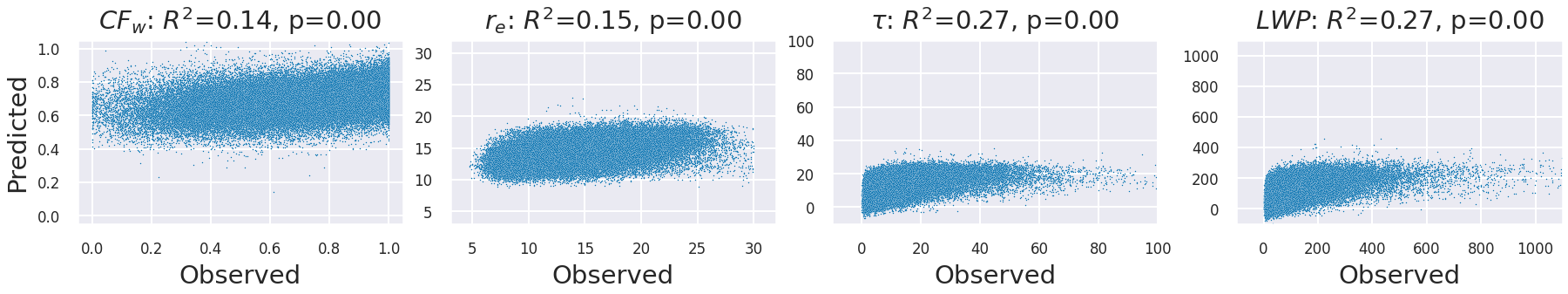}
         \caption{Bayesian Linear Regression}
         \label{fig:fig:scatter_blr}
     \end{subfigure}
     
     \begin{subfigure}[b]{0.95\textwidth}
         \centering
         \includegraphics[width=\textwidth]{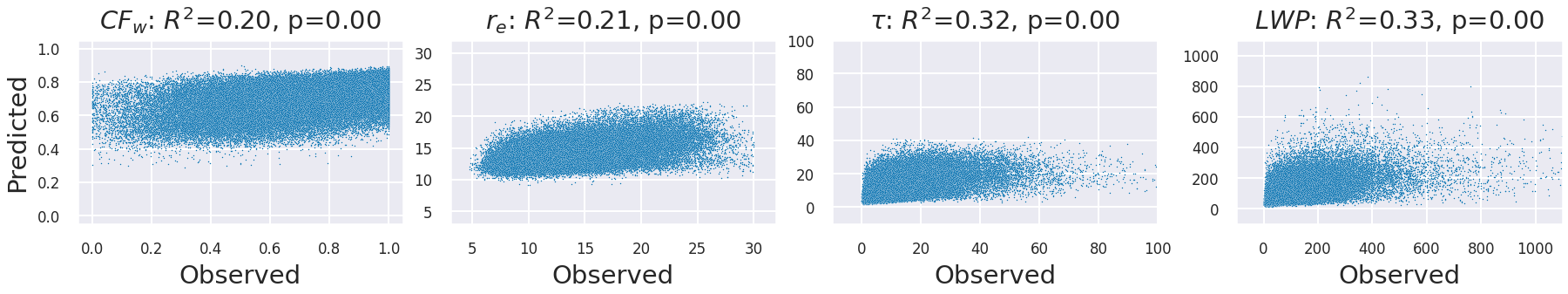}
         \caption{Causal Forest}
         \label{fig:scatter_cf}
     \end{subfigure}
     
     \begin{subfigure}[b]{0.95\textwidth}
         \centering
         \includegraphics[width=\textwidth]{Figures/ensemble/test/scatterplots.png}
         \caption{Quince}
         \label{fig:scatter_quince}
     \end{subfigure}
    \caption{Comparing regression test set accuracy between Quince, Causal Forest and Bayesian Linear Regression. The squared Pearson R coefficient is shown with associated p-value. We see that the non-linear models are better predictors of each outcome.}
    \label{fig:scatter_model_comparison}
\end{figure}

In Figure \ref{fig:scatter_model_comparison}, we compare the results on the test data split for each of the Bayesian Linear Regression, Causal Forest, and Quince methods. We see that the non-linear models improve regression accuracy and a further marginal gain in performance between the Quince and Causal Forest methods.

\begin{figure}[ht]
    \centering
    \begin{subfigure}[b]{0.45\textwidth}
        \centering
        \includegraphics[width=\textwidth]{Figures/ensemble/test/heatmaps_lre.png}
        \caption{Quince: Effect of AOD on $r_e$}
        \label{fig:quince_lre}
    \end{subfigure}
    \begin{subfigure}[b]{0.45\textwidth}
        \centering
        \includegraphics[width=\textwidth]{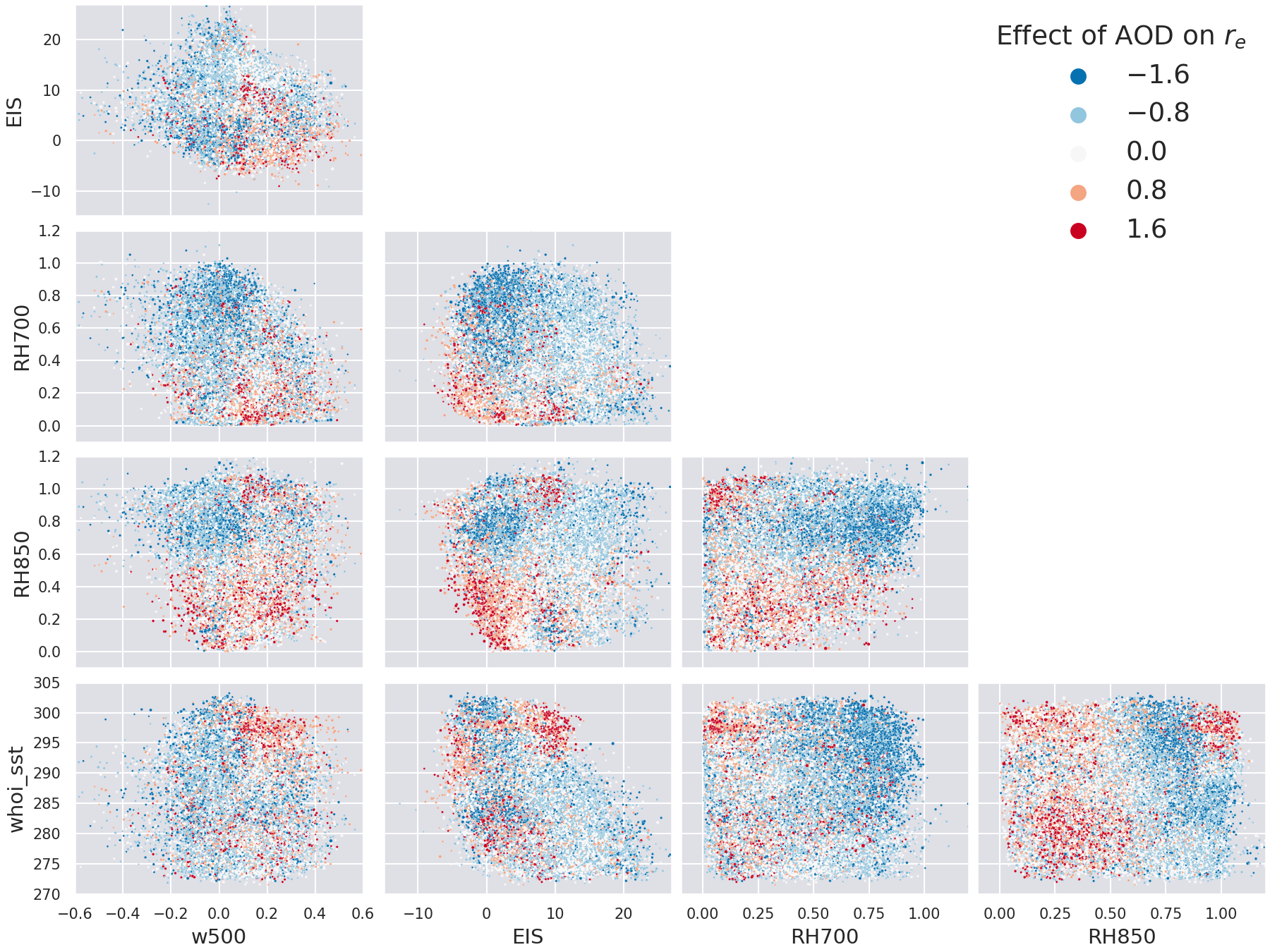}
        \caption{Causal Forest: Effect of AOD on $r_e$}
        \label{fig:rf_lre}
    \end{subfigure}
    
    \begin{subfigure}[b]{0.45\textwidth}
        \centering
        \includegraphics[width=\textwidth]{Figures/ensemble/test/heatmaps_cf.png}
        \caption{Quince: Effect of AOD on $CF_w$}
        \label{fig:quince_cf}
    \end{subfigure}
    \begin{subfigure}[b]{0.45\textwidth}
        \centering
        \includegraphics[width=\textwidth]{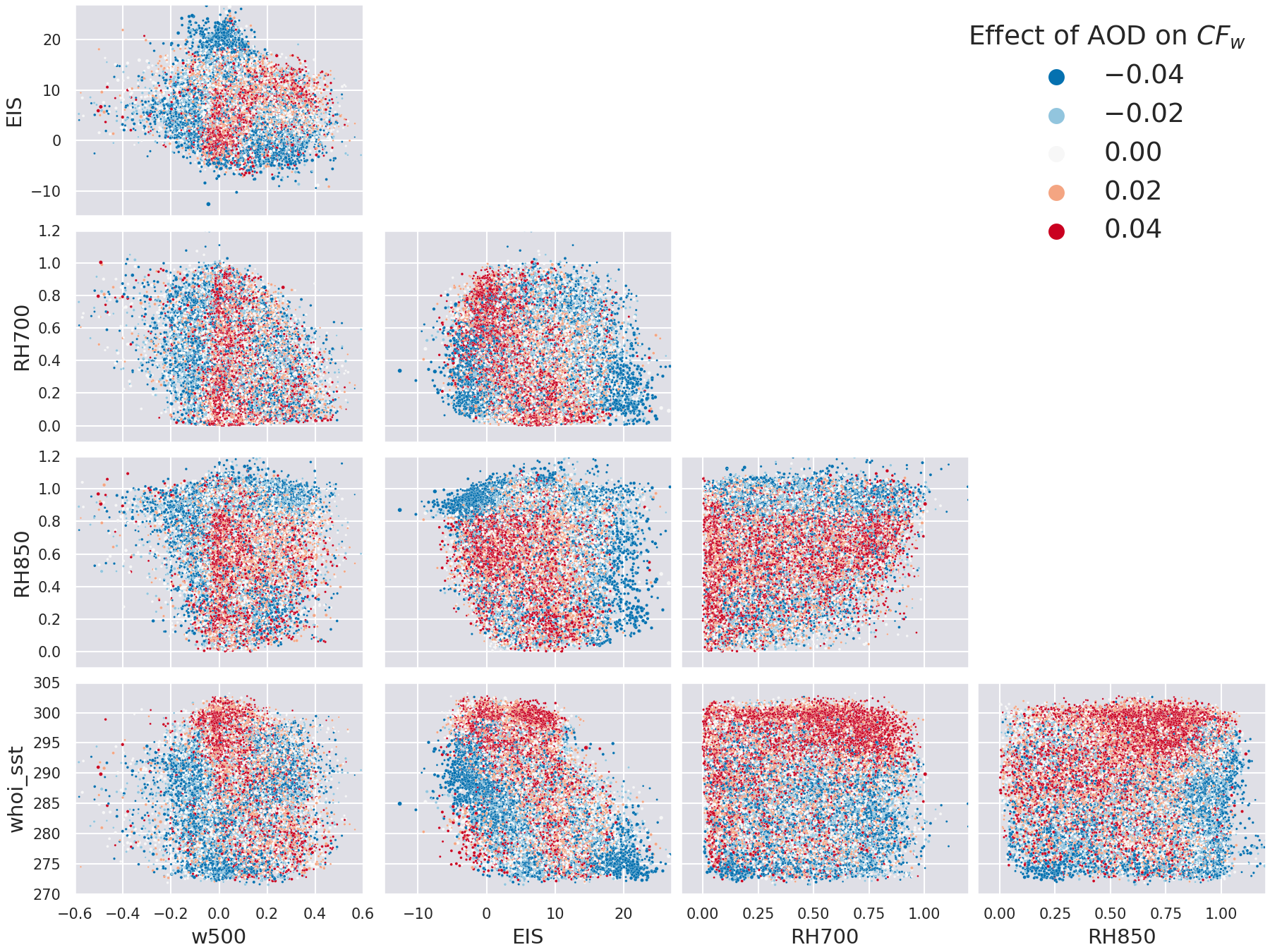}
        \caption{Causal Forest: Effect of AOD on $CF_w$}
        \label{fig:rf_cf}
    \end{subfigure}
    
    \begin{subfigure}[b]{0.45\textwidth}
        \centering
        \includegraphics[width=\textwidth]{Figures/ensemble/test/heatmaps_cod.png}
        \caption{Quince: Effect of AOD on $\tau$}
        \label{fig:quince_cod}
    \end{subfigure}
    \begin{subfigure}[b]{0.45\textwidth}
        \centering
        \includegraphics[width=\textwidth]{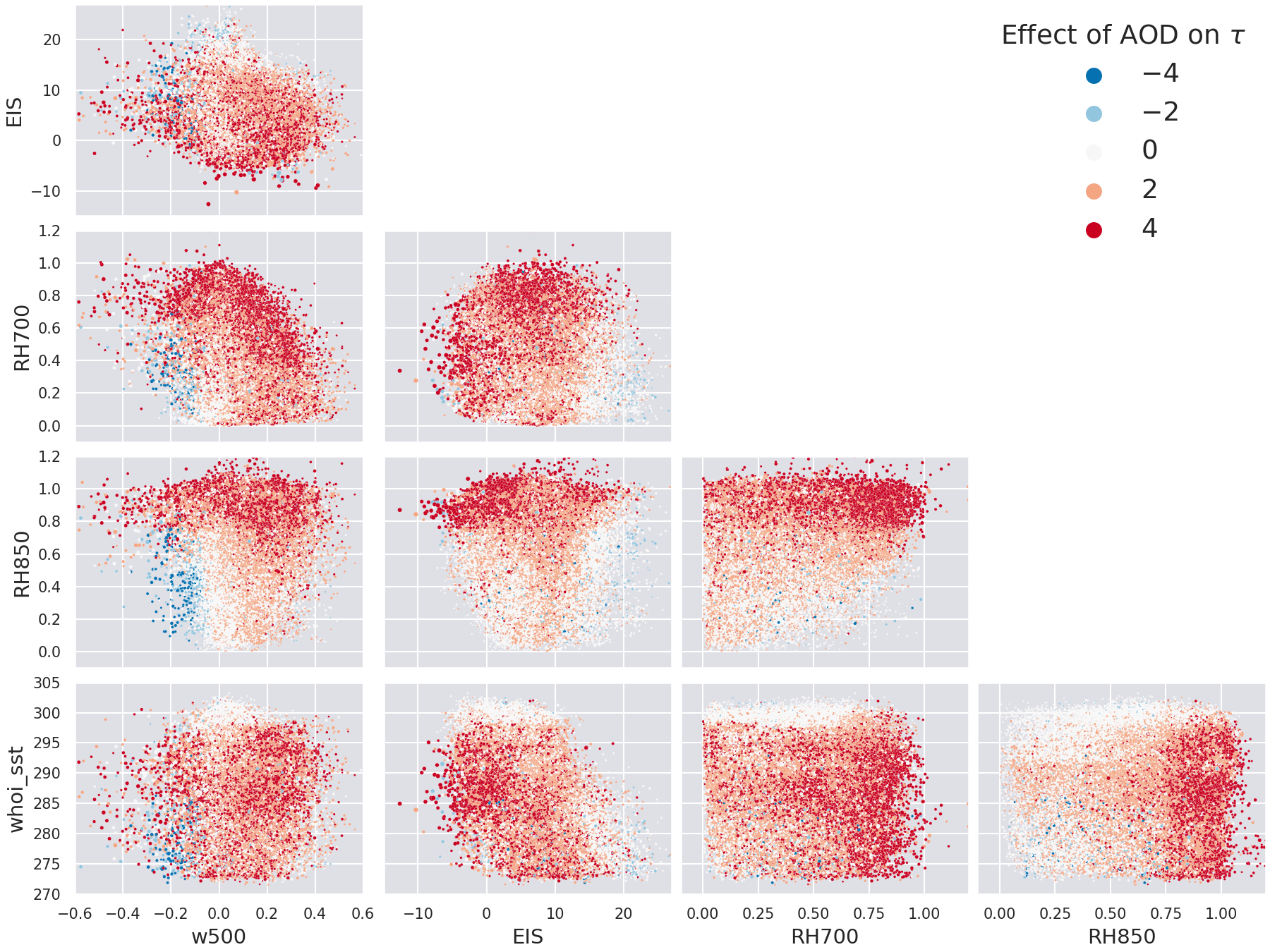}
        \caption{Causal Forest: Effect of AOD on $\tau$}
        \label{fig:rf_cod}
    \end{subfigure}
    
    \begin{subfigure}[b]{0.45\textwidth}
        \centering
        \includegraphics[width=\textwidth]{Figures/ensemble/test/heatmaps_lwp.png}
        \caption{Quince: Effect of AOD on LWP}
        \label{fig:quince_lwp}
    \end{subfigure}
    \begin{subfigure}[b]{0.45\textwidth}
        \centering
        \includegraphics[width=\textwidth]{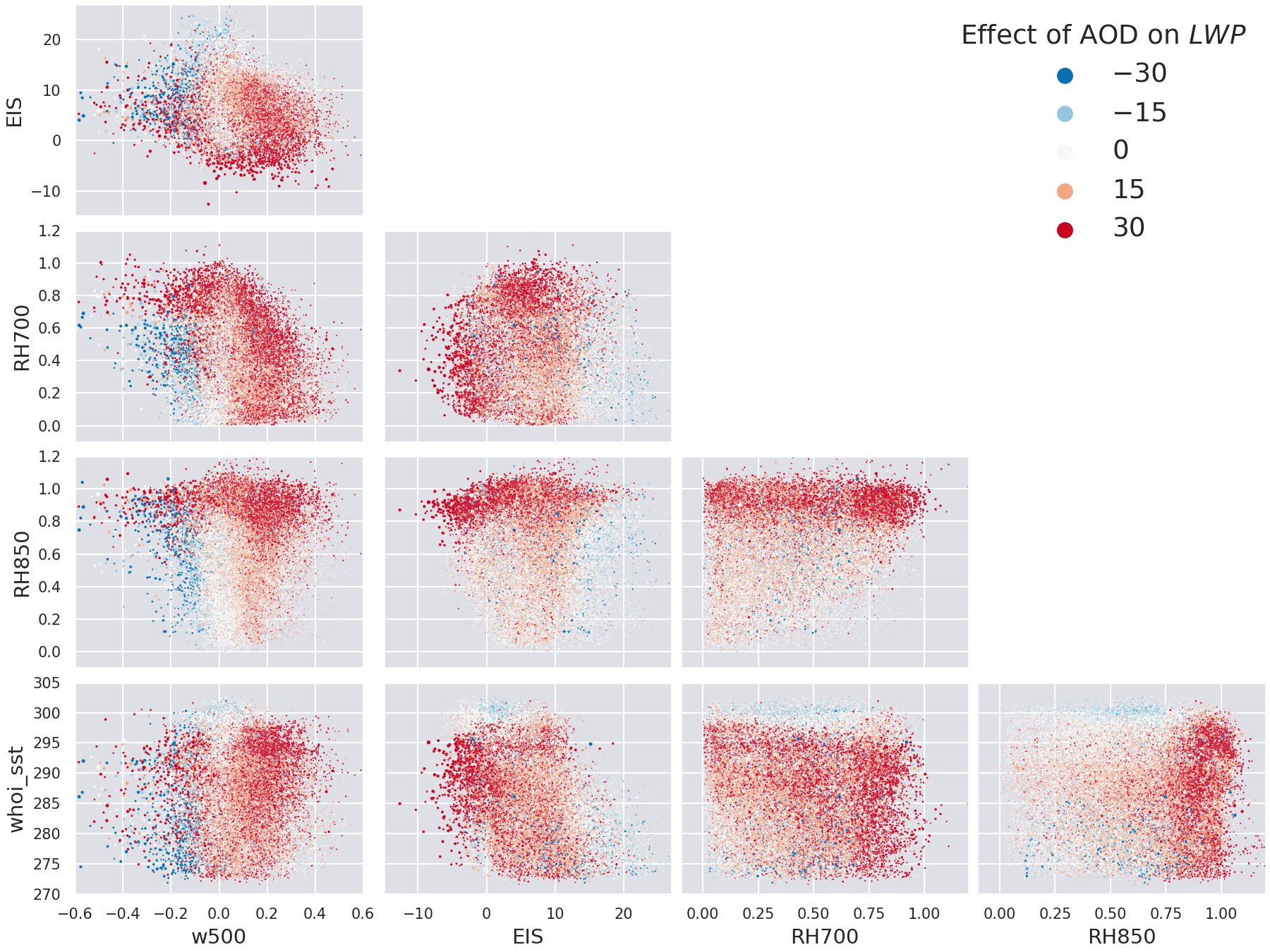}
        \caption{Causal Forest: Effect of AOD on LWP}
        \label{fig:rf_lwp}
    \end{subfigure}
    \caption{Qualitative comparison of heterogeneous effect estimates of AOD on outcome variables between Quince methodology and Causal Forest. We see the same general trends using two different methods which lends evidence to support the treatment effect estimates reported.}
    \label{fig:sanity_rf_quince}
\end{figure}

In Figure \ref{fig:sanity_rf_quince}, we plot the estimated heterogeneous effects from both the Quince and Causal Forest methodologies side by side for each of $CF_w$, $r_e$, $\tau$, and LWP. At a high level we see the same general trends given by both methods.

\begin{figure}[ht]
     \centering
     \begin{subfigure}[b]{0.95\textwidth}
         \centering
         \includegraphics[width=\textwidth]{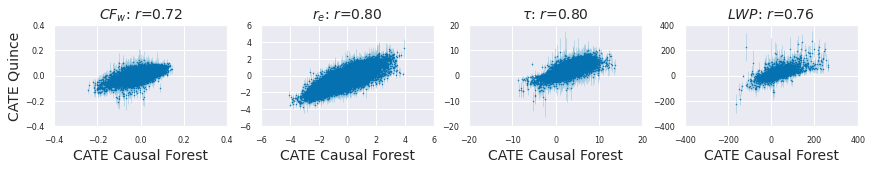}
         \caption{CATE Quince vs. Causal Forest with Quince Uncertainty}
         \label{fig:fig:error_quince}
     \end{subfigure}
     
     \begin{subfigure}[b]{0.95\textwidth}
         \centering
         \includegraphics[width=\textwidth]{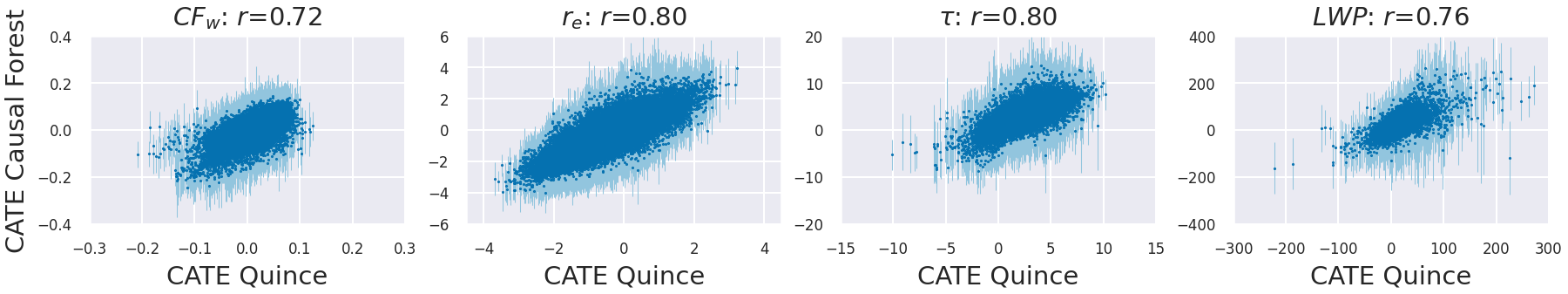}
         \caption{CATE Causal Forest vs. Quince with Causal Forest Uncertainty}
         \label{fig:error_cf}
     \end{subfigure}
    \caption{TODO}
    \label{fig:cate_comparison}
\end{figure}

This general trend is also reflected in Figure \ref{fig:cate_comparison} where we plot the estimated CATE values for the Quince and Causal Forest methods against one another. The Spearman rank correlation coefficient r is also reported. 

However, we can see that the relationship is not perfect and that there is disagreement between the estimated ATE values of each method (Table \ref{tab:ate}). So can we go further to see if we can learn from the discrepancy between predictions?

\vspace{-1em}
\section{Dataset Details}
\vspace{-1em}
All observations of r\textsubscript{e} are from the daily mean, 1\(^{\circ}\) x 1\(^{\circ}\) resolution Cloud Product from MODIS aboard both Aqua and Terra. The effective cloud droplet radius is found for pixels that are both probably or definitely cloudy according to the MODIS Cloud Mask. The NOAA CMORPH CDR Precipitation Product is found by integrating multiple observations of precipitation from both satellite and \textit{in situ} sources. Sea surface temperatures from NOAA WHOI CDR is found using multiple observations of surface brightness temperature and incorporating precipitation estimates in order to better approximate the effects of the diurnal cycle on sea surface temperature. The MERRA-2 model, which calculates global profiles of temperature, relative humidity, and pressure, assimilates hyperspectral and passive microwave satellite observations to enhance its ability to model Earth's atmosphere.

\section{Implementation Details}

\subsection{Quince}

We follow \citet{jesson2020identifying} and use an ensemble of Mixture Density Networks (MDNs) \cite{bishop1994mixture}. Each MDN is adapted for causal-effect inference by following the Dragonnnet architecture of \citet{shi2019adapting}. The deep neural network producing the hidden representation has 3 hidden residual layers with 800 neurons each. Dropout \cite{JMLR:v15:srivastava14a} is applied at a rate of 0.5 after each linear hidden layer followed by ReLU activation functions \cite{fukushima1982neocognitron, nair2010relu}. The treatment prediction head is a linear layer followed by a pytorch Bernoulli distribution layer. Each conditional outcome head is a MDN comprised of a linear layer followed by a pytorch MixtureSameFamily distribution with Normal component distributions. Each MDN has 20 mixure components. The sum of the log likelihoods for both the treatment head and each conditional outcome head multiplied by the observed treatment is minimized using Adam optimization \cite{kingma2017adam} with learning rate of 0.0002 and pytorch default parameters. The model is trained on the training data split for 400 epochs with a batch size of 4096. Model selection is done by evaluating the average R2 score across the 4 outcomes on the validation set and selecting the model parameters at the epoch with the best score. We use an ensemble size of 10.

We use Ray Tune to optimize our hyper-parameters \cite{moritz2018ray, liaw2018tune} with the Bayesian Optimization HyperBand algorithm \cite{falkner2018bohb}. The space we search over is given in Table \ref{tab:hypers}.

\begin{table}[ht]
  \caption{Quince hyper-parameter search space}
  \label{tab:hypers}
  \centering
  \begin{tabular}{lc}
    \toprule
    Hyperparameter & Space \\
    \midrule
    dim hidden & [50, 100, 200, 400, 800] \\
    depth & [2, 3, 4, 5] \\
    num components & [1, 2, 5, 10, 20] \\
    dropout rate & [0.0, 0.1, 0.2, 0.5] \\
    spectral norm \cite{gouk2021regularisation} & [0.0, 0.95, 1.5, 3.0] \\
    negative slope \cite{clevert2015fast, xu2015empirical} & [0.0, 0.1, 0.2, 0.3, elu] \\
    learning rate & [0.0002, 0.0005, 0.001] \\
    batch size & [1024, 2048, 4096] \\
    \bottomrule
  \end{tabular}
\end{table}

\subsection{Causal Forest}

We use the Causal Forest of \citet{athey2019estimating} as implemented in Microsoft's EconML python package \cite{econml}. We use 1000 estimators, max-samples 0.5, discrete-treatment True, and default parameters for the rest. We did a grid search using the validation set over the number of estimators [100, 500, 1000].

\subsection{Bayesian Linear Regression}

We use Bayesian Ridge regression \cite{mackay1992bayesian, tipping2001sparse} from scikit learn with default parameters \cite{scikit-learn}. The model is fit on the training data. We report the coefficient for the treatment input.

\begin{table}[ht]
  \caption{Comparison of the average treatment effect (ATE) estimates of different methodologies. The mean and standard deviation of the approximate posterior distribution (BLR) or mean and standard error of the model ensemble (Causal Forest and Quince) are reported.}
  \label{tab:ate}
  \centering
  \begin{tabular}{lcccccc}
    \toprule
    & \multicolumn{3}{c}{Estimated ATE of AOD on $CF_w$} & \multicolumn{3}{c}{Estimated ATE of AOD on $r_e$} \\
    Method & Train & Valid & Test & Train & Valid & Test \\
    \midrule
    BLR & $-.002\pm.001$ &  &  & $-.29\pm.02$ &  &  \\
    Forest & $.003\pm.030$ & $.005\pm.030$ & $.004\pm.029$  & $-.33\pm.57$ & $-.33\pm.57$ & $-.36\pm.53$ \\
    Quince & $.008\pm.001$ & $.008\pm.001$ & $.008\pm.001$  & $-.43\pm.01$ & $-.41\pm.01$ & $-.45\pm.01$ \\
    Context & $.001\pm.003$ & $.001\pm.003$ & $.001\pm.003$  & $-.06\pm.06$ & $-.04\pm.06$ & $-.10\pm.06$ \\
    \bottomrule \\
    \toprule
    & \multicolumn{3}{c}{Estimated ATE of AOD on $\tau$} & \multicolumn{3}{c}{Estimated ATE of AOD on LWP} \\
    Method & Train & Valid & Test & Train & Valid & Test \\
    \midrule
    BLR & $2.34\pm0.04$ &  &  & $17.8\pm0.40$ &  &  \\
    Forest & $2.05\pm1.37$ & $2.03\pm1.37$ & $2.02\pm1.27$ & $13.3\pm15.6$ & $13.0\pm15.6$ & $12.8\pm14.5$ \\
    Quince & $1.64\pm0.04$ & $1.61\pm0.04$ & $1.61\pm0.04$ & $8.49\pm0.46$ & $8.34\pm0.47$ & $8.05\pm0.46$ \\
    Context & $1.58\pm0.21$ & $1.56\pm0.21$ & $1.52\pm0.22$ & $12.8\pm2.25$ & $12.4\pm2.26$ & $11.7\pm2.30$ \\
    \bottomrule
  \end{tabular}
\end{table}

\end{document}

%% file: commands.tex
\usepackage{amsmath}
\usepackage{amssymb}
\usepackage{amsthm}
\usepackage{mathtools}
\usepackage{bm}
\usepackage{xcolor}
\usepackage{accents}

\definecolor{blue}{RGB}{0,114,178}
\definecolor{red}{RGB}{213,80,20}
\definecolor{green}{RGB}{0,158,115}
\definecolor{purple}{RGB}{204,121,167}
\definecolor{orange}{RGB}{230,159, 0}
\definecolor{pink}{RGB}{204,121,167}

\newcommand\indep{\protect\mathpalette{\protect\independenT}{\perp}}
\def\independenT#1#2{\mathrel{\rlap{$#1#2$}\mkern2mu{#1#2}}}

\newcommand{\E}{\mathop{\mathbb{E}}}

\newcommand{\D}{\mathcal{D}}

\newcommand{\x}{\mathbf{x}}
\newcommand{\X}{\mathbf{X}}

\newcommand{\tf}{\mathrm{t}}

\newcommand{\T}{\mathrm{T}}

\newcommand{\y}{\mathrm{y}}
\newcommand{\Y}{\mathrm{Y}}

\newcommand{\yzero}{\y^0}
\newcommand{\Yzero}{\Y^0}
\newcommand{\yone}{\y^1}
\newcommand{\Yone}{\Y^1}

\newcommand{\yt}{\y^\tf}